\documentclass[aps,prd,twocolumn,showpacs,amsmath,amssymb]{revtex4}

\usepackage{graphicx}
\usepackage{dcolumn}
\usepackage{bm}
\usepackage{epsfig}

\begin{document}

\title{Singularity-free Bianchi spaces with
nonlinear electrodynamics.}

\author{Ricardo Garc\'{\i}a-Salcedo}
\email{rgs@uaemex.mx}
\affiliation{CIESA-FMVZ, U. A. E. Mex., Instituto
Literario 100,
50000, Toluca, M\'exico.}
\author{Nora Bret\'on}
\email{nora@fis.cinvestav.mx}
\affiliation{Departamento de F\'{\i}sica,
Cinvestav--IPN, Apartado Postal 14--740, C.P. 07000, M\'exico,
D.F., MEXICO}

\date{\today}

\begin{abstract}
In this paper we present an analysis to determine the existence of
singularities in spatially homogeneous anisotropic universes filled with
nonlinear electromagnetic radiation. These spaces are conformal to Bianchi
spaces admitting a three parameter group of motions G$_3$.  For these
models we study geodesic completeness.  It is shown that with nonlinear
electromagnetic field some of the Bianchi spaces are geodesically
complete, like G$_3$IX and G$_3$VIII; however, completeness depends on the
curvature of the space. When certain topology is assumed, Bianchi G$_3$IX
presents geodesics that are imprisoned.  It is surprising that in the
linear limit (Maxwell field)  the spacetimes are singularity-free even if
the curvature parameter is zero.
\end{abstract}


\maketitle

\section{Introduction}

General Relativity breaks down at certains points called singularities.
The singularity theorems, that were shown for the first time 30 years ago
by Stephen Hawking, Robert Geroch and Roger Penrose \cite{Hawking}, gave
the bad news that under certain conditions the singularities of the
spacetime would appear in the future or in the past.  An important subject
that is not yet determined is the nature of the singularities of the
spacetime in general relativity.
 
There are several criteria that tell us if a spacetime is singular, such
as energy conditions or the divergence of the curvature invariants.
However, these conditions are only necessary but not sufficient. If one of
them fails to hold, one is ignorant if the spacetime possesses a
singularity or not. The common characteristic of singular spacetimes is
the existence of incomplete causal geodesics. We shall adopt as definition
of the absence of singularities the completeness of causal geodesics
(g-completeness), i. e. no observer in free fall leaves the spacetime in a
finite proper time. We note, however, that not even the geodesic
completeness guarantees the absence of singularities (see a counterexample
by Geroch \cite{Geroch}, of a geodesically complete spacetime with
incomplete non-geodesic curves).

Mars and Senovilla presented a Bianchi IX model with an energy-momentum
tensor of perfect fluid type \cite{Mars} that is completely regular
everywhere in the energy density and presure and is geodesically complete.
This model provides an example of how the presence of matter can help to
avoid the initial singularity while keeping physical compelling dominant
energy conditions.
 
Recently, more studies about singularity-free Bianchi type IX models have
arisen; for example, Berger used the method of consistent potentials
\cite{Berger} to explain how minimally coupled (classical) scalar field
can suppress mixmaster oscilations \cite{Misner} when approaching the
singularity of Bianchi IX cosmology. Toporensky and Ustiansky studied the
possibility of non-singular transition from contraction to expansion in
the dynamics of the Bianchi IX cosmological model with minimally coupled
massive real scalar field \cite{Toporensky}.
          
In this paper we study singularities in spatially homogeneous but
anisotropic spaces where the energy-momentum tensor corresponds to
(source-free) nonlinear electromagnetic field, in particular, Born-Infeld
field. For these spaces the energy momentum tensor satisfies the dominant
energy condition and the matter does not diverge; three cases possess
regular Weyl invariants. These conditions point to the absence of
singularities provided the spaces do pass the test of geodesic
completeness.  The asumed spacetime includes particular cases of Bianchi
spaces. The analysis of the geodesics lead us to conclude that some of the
studied Bianchi spaces avoid the singularity, like Bianchi G$_3$IX
and G$_3$VIII, while Bianchi G$_3$II might present singularity.  
Spaces like Bianchi G$_3$III and G$_3$I as well
as a particular case of Kantowski-Sachs present singularity in the
curvature, i.e. the invariants diverge after a finite time, however, in
two cases the geodesics are complete.

The solutions investigated in this papers have already been discussed.   
Approximate solution was presented in \cite{paper1} and the posibility of
(NLED) as source of inflation was addressed in \cite{paper2}. The
singularity structure is studied in the present paper.

Before getting into the main subject, we give a summary of non-linear
electrodynamics (NLED) in Sec. II. In Sec. III the studied solutions of
the coupled Einstein-Born-Infeld (EBI) equations are presented along
with comments on their geometric structure, energy conditions and scalar
invariants. In Sec. IV we analyze the geodesics and the nature of the
singularities that appear in the models. 
In Sec. V we address the limit of linear electrodynamics
(Einstein-Maxwell) and it turns out that for this field all solutions are
geodesically complete. In the final section we draw some conclusions.

\section{Born-Infeld Nonlinear electrodynamics}

One of the motivations to explore spacetimes with nonlinear
electromagnetic fields is that in early epochs of the universe the
magnetic fields exceeded some $10^{15}$G. At such values of the
electromagnetic field, the interaction of photons with themselves becomes
important and classical electrodynamics is not valid anymore
\cite{Jackson}. To overcome the use of quantum electrodynamics (QED) we
shall employ the Born-Infeld theory.

The nonlinear electrodynamic theory presented by Born and Infeld (1934)  
possesses a Lagrangian which takes into account quantum electrodynamical
corrections, this fact was noted by Euler and Heisenberg soon after the
proposal by Born and Infeld and later on it was remarked by Schwinger. In
the papers by Heisenberg and Euler \cite{Euler} and by Drummond and
Hathrell \cite{Drummond}, it was demonstrated that the Born-Infeld
lagrangian has the same dependence in the electromagnetic invariants than
QED for the one-loop approach. On the other side, based on very general
considerations, Schwinger \cite{Schwinger} obtained an expression for the
lagrangian that coincides, within numerical factors of order unity, with
the BI lagrangian in the same approximation. Therefore, EBI theory should
give us, at least qualitatively, characteristics of early universes that
could not be obtained within classical electrodynamics.
  
Moreover, lately the BI type action has arisen in string theory: it
happens that open strings with Dirichlet boundary conditions (D-branes)
are described by an effective action of Dirac-Born-Infeld, that is related
to the effective action of BI type in the theory of open strings. The
action to low energies corresponding to these objects, called branes, can
be written like a term of BI plus a term of the Weiss-Zumino or
Chern-Simmons type \cite{Tseylin}. The BI strings are solitons that
represent terminal strings on branes.

On the other side, in relation with cosmological models with
electromagnetic fields, Brill \cite{Brill} studied Einstein-Maxwell
solutions in a homogeneous and nonisotropic universe. His solution is a
generalization of Taub's vacuum universe and represents a closed universe
of topology R$\times S^3$ filled with gravitational and electromagnetic
radiation.  This universe expands in two spatial directions and contracts
along the third one; moreover its invariants are finite. Also Hughston and
Jacobs \cite{Jacobs} investigated the behavior of source-free,
homogeneous, electromagnetic fields in Bianchi type cosmologies. They
showed that there can be no pure electric (magnetic) fields in types VIII,
IX, V, VI ($h \ne 0$) and VI ($h \ne -1$). Types III, VI ($h=-1, h=0$)
allow one independent component of the magnetic field. Type II allows two
independent magnetic field components while type I puts no constraints on
the magnetic field components.  Note that results in \cite{Jacobs} were
obtained for Maxwell fields in the Bianchi geometry, non coupled to
Einstein equations.  In a previous paper, Hacyan \cite{Hacyan}
investigated the effect of BI electrodynamics in the initial singularity
of a Bianchi I space. It turned out in that case that the presence of a BI
field smoothed the oscillations of the metric functions when approaching
the singularity.  In the cases addressed in the present paper, Bianchi
types G$_3$I, G$_3$III as well as Kantowski-Sachs admit an electric or
magnetic non linear field (not both) and G$_3$II, G$_3$VIII and G$_3$IX
admit both components, moreover, the presence of the electromagnetic field
prevents the occurrence of singularities in some Bianchi spaces.

\subsection{The Born-Infeld field parameterization}

We describe in what follows the main characteristics of the BI formalism
used to determine the EBI solutions.
  
The Born-Infeld (BI) nonlinear electromagnetic theory is self-consistent
and satisfies all natural requirements. Its lagrangian depends on the BI
constant $b,$ the maximum field strenght; it also depends in nonlinear
form on the electromagnetic invariants, $P$ and $Q$, of the the
antisymmetric tensor $P^{\mu \nu }$, the generalization of the
electromagnetic tensor $F^{\mu \nu }$; the BI lagrangian is given by

\begin{equation}
L_{BI}=-\frac{1}{2}P^{\mu \nu }F_{\mu \nu }+\mathcal{H}(P,Q)  \label{2.1}
\end{equation}
where $\mathcal{H}(P,Q)$ is the so called structural function.  The
structural function is constrained to satisfy some physical requirementes:
(i) the correspondence with the linear theory of Maxwell in the limit when
$b$ goes to infinity $(\mathcal{H}(P,Q)=P+O(P^{2},Q^{2}))$; (ii) the
parity conservation $(\mathcal{H}(P,Q)=\mathcal{H}(P,-Q));$ (iii) the
positive definiteness of the energy density $(\mathcal{H},_{P}>0)$ and,
(iv) the requirement of the timelike nature of the energy flux vector $(P%
\mathcal{H},_{P}+Q\mathcal{H},_{Q}-\mathcal{H}\leq 0).$ Conditions (iii)
and (iv) amount to the fulfilment of the dominant energy condition (DEC)
\cite{Hawking}: $T_{\mu \nu }u^{\mu }u^{\nu }>0$, for every time direction
$ u_{\mu }, (u^{\mu }u_{\mu }<0)$ and $T_{\mu \nu }u^{\mu }$ is a
nonspacelike vector. The energy-momentum tensor in terms of
the structural function is given by:

\begin{equation}
T_{\mu \nu }=-\mathcal{H},_{P}P_{\mu \alpha }P_{\quad \nu }^{\alpha}+
g_{\mu \nu } 
\left( 2P\mathcal{H},_{P}+Q\mathcal{H},_{Q}-\mathcal{H} \right) .
\label{Tmunu}
\end{equation}

The structural function for the BI field is given by 
\begin{equation}
\mathcal{H}(P,Q)=b^{2}-\sqrt{b^{4}-2b^{2}P+Q^{2}},  
\label{structfunc}
\end{equation}
 
In the linear limit, which is obtained by taking $b\rightarrow \infty$,
then $\mathcal{H}$ $=P$ and $P_{\mu \nu }=F_{\mu \nu }$.  $P_{\mu \nu}$
and $F_{\mu \nu }$ are related through the material or constitutive
equations:

\begin{equation}
F_{\mu \nu }=\mathcal{H},_{P}P_{\mu \nu }+\mathcal{H},_{Q}\check{P}_{\mu \nu
},  \label{2.4}
\end{equation}%
where $\check{P}_{\mu \nu }=-\frac{1}{2}{\epsilon}_{\mu \nu \alpha \beta
}P^{\alpha \beta }$ is the dual of $P_{\mu \nu }.$ An extensive treatment
on nonlinear electrodynamics in curved spacetimes can be consulted in
\cite{Pleban}.

In the null tetrad formalism \cite{Plebanski}, and dealing with a Petrov
type-D metric, one can always align the directions of the real null
vectors $e^3$ and $e^4$ along the double Debever-Penrose (DP) vectors. We
shall assume that the eigenvectors of $F_{ab}$ (and consequently the ones
of $P_{ab}$) are also aligned in the direction of the DP vectors. Hence
the novanishing components of $F_{ab}$ are $F_{12}$ and $F_{34}$ ($P_{12}$
and $P_{34}$).  We shall adopt the following parameterization:

\begin{equation}
P_{12}=iH, \qquad P_{34}=D,
\label{parametr}
\end{equation} 
where $D$ is the electric displacement and $H$ is the magnetic field,
the invariants acquire the form:

\begin{equation}
P=\frac{P^{ab}P_{ab}}{4}= - \frac{D^2-H^2}{2}, \quad
Q=\frac{P^{ab} \check{P}_{ab}}{4}=iHD,
\end{equation}

In this parameterization the structural function turns out to be

\begin{eqnarray}
\mathcal{H}&&=b^2- \sqrt{b^4+b^2(D^2-H^2)-H^2D^2} \nonumber\\
&&= b^2- \sqrt{(b^2+D^2)(b^2-H^2)}.
\label{structfunc2}
\end{eqnarray}

The parameterization (\ref{parametr}) has the advantage that there are
only two nonvanishing components of the energy momentum tensor:

\begin{eqnarray}
T_{34}&=&2b^2
[\frac{b^2+D^2}{\sqrt{(b^2+D^2)(b^2-H^2)}}-1]=2b^2(e^{- \nu}-1),
\nonumber\\
T_{12}&=&2b^2
[\frac{b^2-H^2}{\sqrt{(b^2+D^2)(b^2-H^2)}}-1]=2b^2(e^{\nu}-1),
\label{Tmunu1}
\end{eqnarray}

where we have defined:
\begin{equation} 
e^{\nu }=\sqrt{\frac{b^{2}-H^{2}}{b^{2}+D^{2}}}
\label{nu}
\end{equation} 
  
The condition that $ e^{\nu }$ be real imposes the restriction that
$H<b$, i. e. the fields do not reach the maximum allowed field $b$.


\section{Spatially homogeneous spacetimes with non-linear electromagnetic 
field}
  
Since the energy momentum tensor has two nonvanishing eigenvalues,
Eqs. (\ref{Tmunu1}), may be coupled to spacetimes with two preferred
directions, i. e. spacetimes of type D in the classification by Petrov. We
have found solutions to the coupled Einstein-Born-Infeld equations for
spatially homogeneous and anisotropic spaces, among them there are several
cases of Bianchi metrics.

The studied line element is of the form

\begin{equation}
\phi(t) ^{2}ds^{2}= 
-\frac{dt^{2}}{s(t)} +s(t) {(dx +2lz dy)}^{2}
+\frac{dz^{2}}{h(z)} +h(z)dy^{2},
\label{metric}
\end{equation}
where $x$ and $y$ are ignorable coordinates, $h=1 -\epsilon z^{2}$,
with $\epsilon $ and $l$ being constants; $s=s(t)$ and $\phi
=\phi (t)$ are the metric functions that are determined from the
Einstein-Born-Infeld coupled equations.    
Chosing the null tetrad in the form:

\begin{eqnarray}
e^1&=& \bar{e}^2= \frac{1}{\sqrt{2} \phi} \left[ \frac{dz}{ \sqrt{h}}+i
\sqrt{h} dy \right], \nonumber\\
e^3&=& \frac{1}{\sqrt{2} \phi} \left[ \frac{dt}{ \sqrt{s}}+
\sqrt{s}(dx+2lz dy) \right], \nonumber\\
e^4&=& \frac{1}{\sqrt{2} \phi} \left[ - \frac{dt}{ \sqrt{s}}+ 
\sqrt{s}(dx+2lz dy) \right], \nonumber\\
\end{eqnarray}

\begin{table}
\caption{\label{table1}\small{Values of $h(z)$ and $\epsilon$
for which the studied metric is conformal
to the indicated Bianchi types}}
\begin{ruledtabular}
\begin{tabular}{|c|c|c|c|} 
{$\sigma^1$} &{$h(\theta)=1- \epsilon z^2(\theta)$} & $\epsilon$ & Bianchi
type\\
\hline
\hline   
{$dX$}& { $\sin^2{\theta}$} & 1 & K-S \\
{$dX$} & 1 & 0 & G$_3$I \\
{$dX$} & { $\sinh^2{\theta}$} & -1 & G$_3$III \\
{$dX+ 2l\cos{\theta}dy$}& { $\sin^2{\theta}$} &  1 &  G$_3$IX \\
{$dX+ 2l z dy$} & 1 & 0 & G$_3$II \\
{$dX+ 2l\cosh{\theta}dy$} & { $\sinh^2{\theta}$} & -1 & G$_3$VIII \\
\end{tabular}
\end{ruledtabular}
\end{table}
  
The metric function $s(t)$ turns out to be

\begin{equation} 
s(t)={\phi}^{3} \dot{\phi} [C_{1}+ 2b^2
\int{ \frac{(1-e^{\nu})}{\dot{\phi}^2{\phi}^{4}}dt} +
\epsilon \int{\frac{dt}{\dot{\phi}^2{\phi^2}}}],
\label{ssol}
\end{equation} 
where dot simbolizes derivative with respect to time $t$, $C_{1}$ is an
integration constant related to the vacuum case, $b$ is the BI parameter
and $\epsilon$ is the constant curvature parameter. Moreover, $ \phi (t)$
must satisfy the differential equation $\quad \ddot{\phi}+l^{2}\phi =0.$

The system of EBI equations admits two possible solutions for the
electromagnetic function $\nu (t)$, Eq. (\ref{nu}),

\begin{equation} 
e^{\nu(t) }=\sqrt{1\pm \phi^{4}(t)}. 
\end{equation} 
  
Both solutions present regions in which $s(t)>0$ passing then to the
opposite sign, $s(t)<0$. The regions with $s(t)>0$ admit a cosmological
interpretation as spatially homogeneous spaces. For $s(t)<0$ the signature
of (\ref{metric}) changes corresponding to stationary spaces ($t$
direction changes from timelike to spacelike).  We shall study the case
$e^{\nu }=\sqrt{1-\phi ^{4}}.$ The condition $e^{\nu }$ to be real imposes
the restriction that $\phi^{4}<1$ and it determines the range of the
coordinate $t$. The electromagnetic fields are homogeneous and constant on
spacelike hypersurfaces $\{t=$const\}:

\begin{eqnarray}
D(t)&&= \phi^2(t) \cos{[2l \int{\frac{dt}{\sqrt{1-\phi^4}}}]},\nonumber\\
B(t)&&= - \phi^2(t) \sin{[2l \int{\frac{dt}{\sqrt{1-\phi^4}}}]},
\label{DBfields}
\end{eqnarray}

or in terms of the electromagnetic tensor:

\begin{eqnarray}
\phi^2(t) F_{zy}&&= B(t)=e^{\nu}H(t),\nonumber\\
\phi^2(t) F_{tx}&&= E(t)=e^{-\nu}D(t),
\label{fields}
\end{eqnarray}

The results in subsections III.B and III.C were derived in \cite{paper1}
but are included here for completeness.

The line element (\ref{metric}) can be written in terms of a synchronous
time coordinate if we transform $\frac{dt^2}{s(t)}=dT^2$ and with
$\sigma^1=dx+2lz dy$:

\begin{equation}
dS^2=-dT^2+s({\sigma^1})^2+\frac{dz^2}{h} + h dy^2,
\label{syncmetr}
\end{equation}

Written in the form (\ref{syncmetr}) the Bianchi types can be identified
according to Table \ref{table1} (cf. Eqs. (13.1)-(13.2) in \cite{Kramer}).
In case Bianchi G$_3$IX the metric (\ref{syncmetr}) is a particular case
of Taub's universe (cf. Eqs. (6.29) or (8.5) in \cite{Ryan}).
 
\subsection{Bianchi cases included}
 
The spacetime (\ref{metric}) possesses four Killing vectors. The four
generator group G$_4$ has invariant subgroups G$_3$ that can be spatial
rotations or translations (S$_3$ or T$_3$). In particular, we shall pay
attention to the hypersurface-homogeneous spacetimes that can be
identified as Bianchi spaces I, II, III, VIII and IX and one of them as
Kantowski-Sachs.
 
The Killing vectors $X_a$ are classified according to the values of the
curvature parameter, $\epsilon$ that can be $1, 0, -1$:

i) $\epsilon=0$, $h=1$

\begin{eqnarray}
X_1&&= \partial_y, \quad X_2= \partial_x,\nonumber\\
X_3&&= - \partial_z/2l+y \partial_x, \nonumber\\ 
X_4&&= z \partial_y - y \partial_z+l(y^2-z^2)\partial_x.
\label{Ke=0}
\end{eqnarray}

ii) $\epsilon=-1$, $h=1+z^2$,

\begin{eqnarray}
X_1&&= \partial_y, \quad X_2= \partial_x,\nonumber\\
X_3&&= \sqrt{h} \cosh{y}\partial_z-\frac{\sinh{y}}{\sqrt{h}}(2ly
\partial_x+z \partial_y), \nonumber\\
X_4&&= \sqrt{h} \sinh{y}\partial_z-\frac{\cosh{y}}{\sqrt{h}}(2ly
\partial_x+z \partial_y),
\label{Ke=-1} 
\end{eqnarray}

iii)$\epsilon=1$, $h=1-z^2$,

\begin{eqnarray}
X_1&&= \partial_y, \quad X_2= \partial_x,\nonumber\\
X_3&&= \sqrt{h} \cos{y}\partial_z-\frac{\sin{y}}{\sqrt{h}}(2ly 
\partial_x-z \partial_y) , \nonumber\\
X_4&&= \sqrt{h} \sin{y}\partial_z+\frac{\cos{y}}{\sqrt{h}}(2ly
\partial_x-z \partial_y)
\label{Ke=1}
\end{eqnarray}

Special cases of Bianchi metrics with a particular topology can be
obtained as follows:

i)$\epsilon=0$

In the metric (\ref{metric}) with $\epsilon=0$
we obtain $h(z)=1$ and the line element is   

\begin{equation}
\phi(t) ^{2}ds^{2}=
-\frac{dt^{2}}{s(t)} +s(t) {(dx +2lz dy)}^{2}+dz^{2} +dy^{2},
\label{metricII}
\end{equation}  
where the ranges of all the coordinates are from $-\infty$ to $\infty$,
this is a non-compact spacetime with topology $R^4$ that corresponds to
Bianchi II. If $l=0$, the metric becomes diagonal and can be identified
as a Bianchi I,

\begin{equation}
\phi(t) ^{2}ds^{2}=
-\frac{dt^{2}}{s(t)} +s(t) {dx}^{2}+dz^{2} +dy^{2}.
\label{metricI}
\end{equation}

ii)$\epsilon=-1$

In the metric (\ref{metric}) we transform $z \to \sinh{\theta}$  
obtaining $h(\theta)=\cosh^2{\theta}$ and the line element becomes

\begin{eqnarray}
\phi(t) ^{2}ds^{2}=&&
-\frac{dt^{2}}{s(t)} +s(t) {(dx +2l\sinh{\theta} dy)}^{2}+ \nonumber\\
&&d{\theta}^{2} + \cosh^2{\theta} dy^{2},
\label{metricVIII}
\end{eqnarray}
where the ranges of all the coordinates are from $-\infty$ to $\infty$,
this is a non-compact spacetime with topology $R^4$ that corresponds to
Bianchi VIII. If $l=0$ in the metric (\ref{metric}) and we transform $z
\to \cosh{\theta}$, $h(\theta)=\sinh^2{\theta}$, then the line element
is

\begin{equation}
\phi(t) ^{2}ds^{2}=
-\frac{dt^{2}}{s(t)} +s(t) {dx}^{2} + d{\theta}^{2}
+\sinh^2{\theta} dy^{2},
\label{metricIII}  
\end{equation}
that can be identified as a Bianchi III space.

iii)$\epsilon=1$

In the metric (\ref{metric}) we transform $z \to \sin{\theta}$ and
we obtain $h(\theta)=\cos^2{\theta}$ and the line element as

\begin{equation}
\phi(t) ^{2}ds^{2}=
-\frac{dt^{2}}{s(t)} +s(t) {(dx +2l\sin{\theta} dy)}^{2}+d{\theta}^{2}
+\cos^2{\theta} dy^{2},
\label{metricIX}
\end{equation}
where the ranges of the coordinates are $-\infty < t <\infty$, $0 \le
\theta \le \pi$; if $0 \le y \le 2 \pi$, $0 \le x \le 4 \pi$ it is a
compact spacetime with topology $R \times S^3$ that corresponds to Bianchi
IX. For $l=0$ in metric (\ref{metric}) and changing coordinates to $z \to
\cos{\theta}$ and with $\epsilon=1$ we obtain $h(\theta)=\sin^2{\theta}$
and the line element transforms into

\begin{equation}
\phi(t) ^{2}ds^{2}=
-\frac{dt^{2}}{s(t)} +s(t) {dx}^{2}+d{\theta}^{2}
+\sin^2{\theta} dy^{2},
\label{metricK-S}
\end{equation}
it is a spatially-homogeneous solution which admits no simply-transitive
G$_3$ that can be identified as a Kantowski-Sachs space.

\subsection{Energy conditions}

According to Raychaudhuri, the equation that goberns the volume
expansion of geodetic congruences, $\theta$, is

\begin{equation}
\frac{d \theta}{d \lambda}= -R_{ab}V^{a}V^{b}+2 \omega^2-2
\sigma^2-\frac{\theta^2}{3}+\dot {V}^{\alpha}_{; \alpha}
\label{Ray}
\end{equation}
where $V^{\alpha}$ is a tangent vector to the geodesics, $R_{ab}$ is the
Ricci tensor, $\omega$ is the vorticity, $\sigma$ is the shear and $\dot
{V}^{\alpha}$ is the acceleration of geodesics (or flow lines in a
fluid); we shall consider that $\omega=0$.
   
From Eq. (\ref{Ray}) it can be seen that the expansion $\theta$ of a
timelike geodesic congruence with zero vorticity will monotonically
decrease along a geodesic if, for any timelike vector $V^a$,
$R_{ab}V^{a}V^{b} \ge 0$. This is the strong energy condition (SEC), as
stated in \cite{SenoGRG}, it says: {\it ``A spacetime satisfies the strong
energy condition if $R_{ab}V^{a}V^{b} \ge 0$, for all causal vectors
$V^a$"}.  The term involving the Ricci tensor, $R_{ab}$, in Eq.
(\ref{Ray}) induces contraction of the geodesic lines, indicating that the
focusing of neighbouring geodesics requires SEC.

For our Bianchi spaces the two nonvanishing components of the
traceless Ricci tensor, in terms of the matter content, are

\begin{equation}
R_{12}=2b^2(1-e^{\nu}), \quad R_{34}=2b^2(1-e^{- \nu}).
\end{equation}

If we consider the congruence $V^{a}=\phi \sqrt{s} \partial_t$, or in the
null tetrad formalism, $V^3=V_4=\frac{1}{\sqrt{2}}, V^1=V^2=0$, for the BI
field SEC amounts to $R_{ab}V^{a}V^{b}=R_{34}=2b^2(1- e^{- \nu}) < 0$,
that clearly violates SEC.

The violation of SEC is not exceptional, we remind that minimally coupled
scalar field violates SEC and indeed curvature-coupled scalar field theory
also violates SEC. This point has been discussed recently in
\cite{SenoGRG} and \cite{Visser}. The aspect of interest for us is that
the violation of this condition could imply the absence of an initial
singularity since the fulfiling of SEC is related with the convergence of
neighbouring geodesics.

On the other side, the dominant energy condition (DEC) that states {\it
``for every timelike vector, $V^a$, $T_{ab}V^{a}V^{b} \ge 0$ and
$T_{ab}V^{a}$ is a nonspacelike vector"} \cite{Hawking}, is indeed
fulfilled by the BI energy-momentum tensor, Eqs.
(\ref{Tmunu})-(\ref{structfunc}). DEC means that to any observer the local
energy density appears nonnegative and the local energy flow vector is
non-spacelike. In the BI case DEC is a requirement of a physically
reasonable nonlinear structural function $\mathcal{H}$ and its fulfilment
is guaranteed by conditions (iii) and (iv) as stated in Sec. II.A.

\subsection{Scalar Polynomial Invariants}

For these solutions the only nonvanishing Weyl scalar is given by  

\begin{equation}
\Psi _{2}=\frac{\phi ^{2}}{12}\left( \ddot{s}-8sl^{2}+2\epsilon
+6il \dot{s} \right).
\label{Weylscalar}
\end{equation}

The scalar polynomial invariants constructed from $\Psi _{2}$ are given by
$I=3\Psi _{2}^{2}$ and $J=-\Psi_{2}^{3},$ in such a manner that if $\Psi
_{2}$ is unbounded then both $I$ and $J$ must be unbounded and a curvature
singularity occurs. The regularity and boundedness of $\Psi _{2}$ depends
on the proper for $\ddot{s}(t), \dot{s}(t)$ and $s(t)$. In case of a
suspected singular point, we must check if $\Psi _{2}$ is regular when
evaluated at the suspicious point. If $\Psi _{2}$ remains bounded then it
proceeds to investigate if the geodesics are complete.
 
One could think that regularity of the curvature invariants might be
helpful to avoid singularities but there are known examples of spacetimes
with regular invariants, such as Taub-NUT, that enclose geodesics that
are not complete in their affine parameterization due to the phenomenon
called {\it imprisoned incompleteness} \cite{Hawking}, in which a
non-spacelike curve as it follows to its future enters and remains within
a compact set. In spite that Hawking himself asserted that {\it ``the kind
of behavior which occurs in Taub-NUT space cannot happen if there is some
matter present"}, when studying a spacetime with regular invariants we
must show that in fact geodesics are complete if one wants to assert that
the space is singularity-free.


\begin{figure}\centering
\epsfig{file=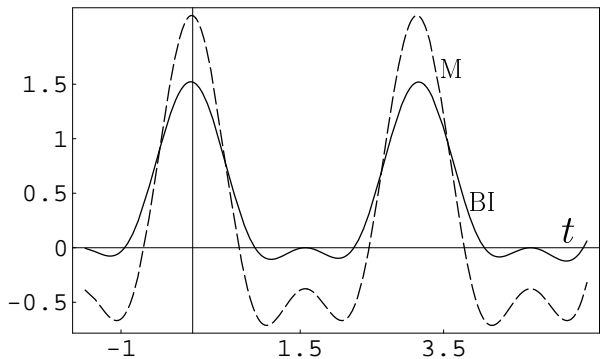, width=8cm}
\caption{
The metric functions $s(t)$ when the BI field is present (continuous
curve) and in the linear limit (Maxwell field) (dashed curve) are shown.  
In general, the presence of BI field smooths oscillations of the metric
function. The plot corresponds to Bianchi G$_3$ IX, the values of the
constants are $l=1$ $A=0.8, C_1=0.5, b=0.85, \epsilon= 1$ and the value
of the constant linear field is $C=0.85$. }
\label{BI-Max}
\end{figure}


\subsection{Linear limit}

The linear (Maxwell) limit can be obtained from the solution
Eq. (\ref{ssol}) taking the limit $ b \to \infty$, obtaining

\begin{equation}
s(t)={\phi}^{3} \dot{\phi} [C_{1}+ C^2
\int{ \frac{dt}{\dot{\phi}^2{\phi}^{4}}} +
\epsilon \int{\frac{dt}{\dot{\phi}^2{\phi^2}}}],
\label{sMax}
\end{equation} 
where $C^2=H^2+D^2$ is related to the energy density of the field;
in this case the electromagnetic field is given by

\begin{equation}
H= C \sin{(2lat)}, \quad D= C \cos{(2lat)},
\label{linearfield}
\end{equation}
where $a$ is a constant. In Fig. \ref{BI-Max} are displayed both metric
functions, the linear (Maxwell)  and Born-Infeld, the values of the
constants correspond to a case of Bianchi G$_3$ IX. In general the BI
field smooths oscillations of the metric function $s(t)$. It is surprising
that with linear electromagnetic field as matter the spacetimes are
singularity-free. We shall address geodesic completeness for this case in
Sec. V.
 

\section{Analysis of singularity structure}
    
Hawking and Ellis \cite{Hawking} discussed Bianchi I spaces and gave a
theorem asserting that singularities will occur in all non-empty spatially
homogeneous models in which the timelike convergence condition
($R_{ab}u^{a}u^{b} \ge 0$) is satisfied. However, as we showed in section
III.B, in the studied case the timelike convergence condition is not
fulfilled. Therefore the possibility of the absence of singularity
persists.
      
For our Bianchi spaces with nonlinear electrodynamics there are two
branches characterized by the value of the constant $l$ that appears in
the non-diagonal term of the metric, $(dx+2lz dy)$.  In the case $l=0$ the
conformal function $\phi(t)$ fulfills the equation $\ddot \phi=0$, then
$\phi = At+ B$ where $A$ and $B$ are constants. In the case $l \ne 0$ the
equation for $\phi(t)$ is $\ddot \phi + l^2 \phi=0$, with solutions $\phi
= A \cos{(lt+B)}$, $A$ and $B$ being constants. The analytical expression
of the metric function $s(t)$ depends on the conformal function $\phi(t)$
and its behavior is different in each case.

We will consider a spacetime as singularity-free when it is geodesically
causal complete (g-completeness), that means intuitively that an observer
in free falling does not leave the spacetime in a finite proper time;
equivalently, that every geodesic can be extended to arbitrary values of
its affine parameter. The demonstration of geodesic completeness is not
simple in most cases \cite{SenoGRG}. Nevertheless, the analysis of the
geodesic extension can be simplified when there exist constants of motion
associated to Killing vectors.
   
The method that we use to analize the completeness is to obtain the first
order equations for the derivatives of the coordinates with respect to an
affine parameter, ${x}_{,\tau}$, ${t}_{,\tau}$, ${y}_{,\tau}$,
${z}_{,\tau}$, using the constants of motion. Then showing that these
first derivatives are bounded, this implies that the corresponding
geodesic curves ${x}(\tau)$, ${t}(\tau)$, ${y}(\tau)$, ${z}(\tau)$, are
complete \cite{Arnold}.

We shall analyze geodesic completeness of the spacetimes in the
following subsections, first for the class of non-diagonal metrics, $l
\ne 0$ that includes Bianchi G$_3$II, G$_3$VIII and G$_3$IX and then in
the next subsection for the diagonal metrics Kantwoski-Sachs and Bianchi
spaces G$_{3}$I and G$_{3}$III.

\subsection{Bianchi  G$_3$II, G$_3$VIII and  G$_3$IX}
   
In the case that $l \ne 0$ the equation for $\phi$ is $\ddot{\phi}+l^2
\phi=0$, with solution $\phi =A \cos{(lt + B)}$, $A, B$ being constants;
the sector $(t,x)$ of the metric is non diagonal and the cases included
are Bianchi G$_3$VIII and G$_3$IX for $\epsilon=-1, 1$, respectively and
G$_3$II when $\epsilon=0$.

The metric function $s(t)$, Eq. (\ref{ssol}) can be integrated in exact
form; with $\phi= A \cos{t}$ (taking $l=1$ and $B=0$) we obtain a
cumbersome expression in terms of $\sin{t}$, $\cos{t}$ and elliptic
functions,
 
\begin{widetext}
\begin{eqnarray} 
s(t)&&= -A^4C_1 \cos^3{t} \sin{t}
+ \epsilon ( \cos^4{t}-\cos^2{t} \sin^2{t}) +\nonumber\\
&&\frac{2b^2}{A^2} \{ \cos^4{t} -\frac{5}{3}\cos^2{t}
\sin^2{t}-\frac{1}{3}\sin^2{t}-
\frac{1}{6}(2\cos{2t}+\cos{4t})
\sqrt{(2-A^2-A^2\cos^2{t})(2+A^2+A^2\cos^2{t})}+\nonumber\\
&&\frac{8}{3}A^2\cos^4{t} \sin^2{t}
\sqrt{\frac{(2+A^2+A^2\cos^2{t})}{(2-A^2-A^2\cos^2{t})}}-
\frac{2}{3} \sin{t} \cos^3{t}[ \sqrt{1+A^2} {\rm E}[ \alpha,
\beta] -\left( \frac{4-A^4}{\sqrt{1-A^2}}\right) {\rm F}[ \gamma, \beta]]
\},\nonumber\\
\alpha&=& \arcsin \left({\frac{\sqrt{2} \sin{t}}{\sqrt{2-A^2-A^2
\cos{2t}}}}\right),
\quad \beta=\frac{2A^2}{1+A^2}, \quad
\gamma={\rm arcsinh} \left({\frac{(1-A^2) \sqrt{2+A^2+A^2\cos{2t}}}{2
\sin{t}}}\right),
\label{sexact}
\end{eqnarray}
\end{widetext}  
 
E and F are the elliptic integrals of second and first kind, respectively.
From the expression (\ref{sexact}), it is clear that $s(t)$ is a periodic
and bounded function. There are values of $t$ for which the metric
function $s(t)$ is null and there the metric (\ref{metric}) is singular.  
For $\phi(t)= A \cos(t)$, these values are $t=\pm (2n+1) \pi /2, n=0, 1,
2,...$ and there we might have a singularity.
There are additional zeroes when negative terms cancel the
positive ones, however, the position of these zeroes depends on the
balance between the distinct parameters, $C_1, b^2/A^2, A^2$ and
$\epsilon$.


\begin{figure}\centering
\epsfig{file=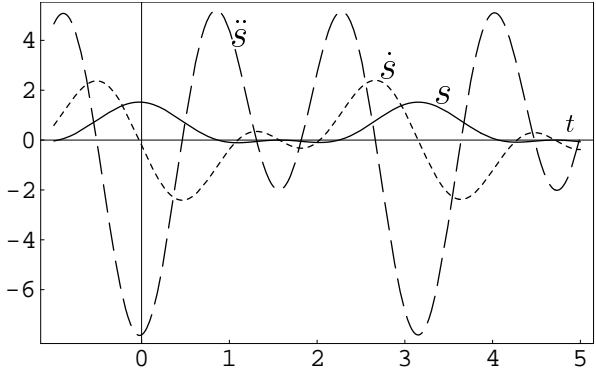, width=8cm}
\caption{
This is the characteristic plot for the metric function $s(t)$ and its
derivatives, $\dot{s}(t), \ddot{s}(t)$ in the case $l \ne 0$. The three
functions are periodic and bounded.  The corresponding Weyl scalar,
$\Psi_2$ does not diverge in the full range $ - \infty < t < \infty$. In
this graphic the values of the constants are $A=0.8, C_1=0.5, b=0.85$ and
$\epsilon= 1$. For $\epsilon=0,-1$ the behavior is qualitatively the same.
}
\label{snon-diag}
\end{figure}

 
The characteristic behavior of the functions $s(t),
\dot{s}(t),\ddot{s}(t)$ is illustrated in Fig.\ref{snon-diag}. The three
functions $s(t), \dot{s}(t), \ddot{s}(t)$ are periodic and bounded, that
do not diverge over all the range $-{\infty}< t < \infty$. As a
consequence, the Weyl scalar $\Psi_2$, Eq. (\ref{Weylscalar}), is also
finite over all the range $-{\infty}< t < \infty$ and so are its derived
invariants. To dilucidate the behavor of the spacetime at $t= \pi/2$ we
shall invetigate the completeness of geodesics.


For each Killing vector $X_i$ there exists a conserved quantity
$x^{\mu}_{,\tau}X_{\mu}=x^{\mu}_{,\tau}g_{\mu \nu}X^{\nu}= $const, where
$x^{\mu}(\tau)$ is a geodesic parametrized by $\tau$.  Since
$\partial_{x}$ and $\partial_{y}$ are Killing vectors, it implies that
there are constants $K_1, K_2$ given by

\begin{eqnarray}
K_1&=&g_{xx}x_{,\tau}+g_{xy}y_{,\tau}, \nonumber\\
K_2&=&g_{yy}y_{,\tau}+g_{yx}x_{,\tau},
\label{KL}
\end{eqnarray}
 
From Eqs. (\ref{KL}) we determine $x_{,\tau}$ and $y_{,\tau}$:

\begin{eqnarray}
y_{,\tau}&=& \frac{\phi^2 }{h}(K_2-MK_1),\nonumber\\
x_{,\tau}&=& \frac{\phi^2 K_1}{s}-My_{,\tau},
\label{x,y}
\end{eqnarray}
where $M=2lz$.  From Eqs. (\ref{x,y}) we can analyze completeness for
$x(\tau)$ and $y(\tau)$. All the coefficients in Eqs. (\ref{x,y}) are
finite ($h=1- \epsilon z$) except for the first term in $x_{,\tau}$, since
there will be divergences if $s/ \phi^2$ is null. We shall analyze this
term with detail.

Taking the limit $t \to \pi/2$,

\begin{equation}
\lim_{t \to \pi/2} \left(\frac{s}{\phi^2} \right)= -\epsilon,
\end{equation}
therefore, only in the case that $\epsilon=0$, the geodesics $x(\tau)$ can
be incomplete. This case corresponds to spaces Bianchi II. Even in this   
case we can have complete geodesics when the constant $K_1=0$, that means,

\begin{equation}
K_1= \frac{s}{\phi^2}(x_{,\tau}+ M y_{,\tau})=0,
\end{equation}
in that case the geodesics $x_{,\tau}=- M y_{,\tau}$, but the equation for
$y_{,\tau}$ is given in terms of bounded coefficients, also $M=2lz$ is
bounded, then in this particular case of $\epsilon=0$, the geodesics
$x(\tau)$ are complete. However the generic behavior when $\epsilon=0$ is
geodesically uncomplete.

To obtain the equations for $t_{,\tau}$ and $z_{,\tau}$ we shall use the
rest of the Killing vectors, quoted in Sec. III.A. These Killing vectors
are different depending on the value of the curvature parameter
$\epsilon$. We separate in three cases: $\epsilon = 1, -1, 0$.
 
For $\epsilon=1$ the Kiling vectors Eqs. (\ref{Ke=1}) imply the existence
of two conserved quantities, $K_3$ and $K_4$ that permit determine
$z_{,\tau}$ as

\begin{equation}
z_{,\tau}^2+\phi^4 (zK_2-2lK_1)^2- \phi^4 h (K_3^2+K_4^2)=0,
\label{z,1}
\end{equation}

On the other hand, we can also use the line element, to 
obtain $t_{,\tau}$,

\begin{equation}
ds^2= \delta= \frac{z_{,\tau}^2}{\phi^2 h}+\frac{h}{\phi^2}
y_{,\tau}^2-\frac{t_{,\tau}^2}{\phi^2 s}+ \frac{s}{\phi^2}(
x_{,\tau}+My_{,\tau})^2,
\end{equation}
where $\delta$ is 1 for spacelike geodesics, 0 for null and -1 if the
geodesics are timelike.

Substituting the constants of motion we have,

\begin{eqnarray}
t_{,\tau}^2&=&s \phi^4 \{ (K_3^2+K_4^2)- (zK_2-2lK_1)^2/h+\nonumber\\
&&(K_2-K_1M)^2/h \}+\phi^4K_1^2- \phi^2 s \delta.
\end{eqnarray}

In the equations for $t_{,\tau}^2$ and $z_{,\tau}^2$ all the coefficients
are bounded, then the geodesic curves $t(\tau)$ and $z(\tau)$ are
complete.

In an analogous way can be determined $t_{,\tau}$ and $z_{,\tau}$ for the
case $\epsilon =-1$. Using the Killing vectors $X_3$ and $X_4$, Eqs.
(\ref{Ke=-1}) we obtain:

\begin{eqnarray} 
z_{,\tau}^2&=& \phi^4 (zK_2+2lK_1)^2+ \phi^4 h (K_4^2-K_3^2),\nonumber\\
t_{,\tau}^2&=&s \phi^4 \{ (K_4^2-K_3^2)+ (zK_2+2lK_1)^2/h+\nonumber\\
&&(K_2-K_1M)^2/h \}+\phi^4K_1^2- \phi^2 s \delta.
\end{eqnarray}

For $\epsilon =0$ we have from the motion constants that imply
the Killing vectors in Eqs. (\ref{Ke=0}), 

\begin{eqnarray}
z_{,\tau}^2&=& 4 l \phi^4 h^2 
\{lK_3^2-K_4K_1-K_1(lK_1z^2-K_2z)\},\nonumber\\
t_{,\tau}^2&=&s \phi^4 \{ 4l^2h(K_3-K_1y)^2+
(K_2-K_1M)^2/h \}+\nonumber\\
&&\phi^4K_1^2- \phi^2 s \delta.
\label{z,t,e=0}
\end{eqnarray}
 
In the three cases, $\epsilon=1,-1,0$ the geodesics $t(\tau)$ and
$z(\tau)$ are complete.
 
      
If one considers the topology $RXS^3$ for Bianchi G$_3$IX (case
$\epsilon=1$) then $0<x<4 \pi$ and $x$ reaches a finite value as $\tau$
increases. This geodesic ``wraps" around the $x$-direction an infinite
number of times. It corresponds to the ``phenomena" known as imprisoned
incompleteness, described by Hawking (see paragraph 8.5 in \cite{Hawking})
for Taub-NUT spaces. This pathological behavior arises essentially because
$x$ is identified at $0$ and $4 \pi$, a consequence of the compactness of
Bianchi type IX.

However both types IX and VIII can be considered non-compact and we can
take the range of $x$ being $(- \infty, \infty)$. Being $x$ no longer
periodic the geodesics $x(\tau)$ can be extended indefinitely and the
spaces are complete. The space VIII resembles the ones studied by Siklos
\cite{Siklos} as extensions of Taub-NUT metrics.

\subsection{Kantowski-Sachs and Bianchi spaces $G_{3}I$ and $G_{3}III$}

Let us address the case $l=0$. The included spaces are
Kantowski-Sachs and Bianchi spaces $G_{3}I$ and $G_{3}III$
for $\epsilon= 1,0,-1$, respectively. 


\begin{figure}\centering
\epsfig{file=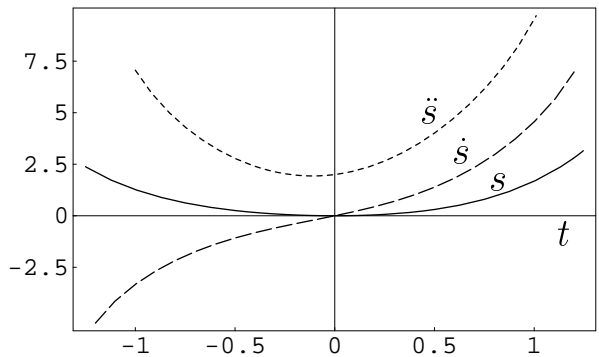, width=8cm}
\caption{ 
It is shown the plot for the metric function $s(t)$ and its derivatives,
$\dot{s}(t), \ddot{s}(t)$ for the case $l=0$. The three functions are
bounded at $t=0$, but $\ddot{s}(t)$diverges at $at \to 1$. In this graphic
the values of the constants are $A=0.8, C_1=0.5, b=0.8$ and $\epsilon=1$.
For $\epsilon=0,-1$ the behavior changes quantitatively.}
\label{sdiag}
\end{figure} 


In this case $\phi=At$ is only a scaling of time coordinate;
$s(t)$, Eq.(\ref{ssol}), can be integrated in exact form as

\begin{eqnarray}
s(t)&=&
\frac{2\sqrt{2}}{3}Ab^2t^3 
\{ {\rm{K}[\frac{1}{2}]}- {\rm{F}}[ \frac{\pi}{2} -
\arcsin{At},\frac{1}{2}] \}+ \nonumber\\
&& \frac{2b^2}{3A^2}( \sqrt{1-A^4t^4}-1) + C_1A^4 t^3 -\epsilon t^2,
\label{s-diag}
\end{eqnarray}
where F is the elliptic integral of the first kind and K is the complete
elliptic integral of the first kind. We observe from Eq. (\ref{s-diag})
that $s(t)$ take real values in the interval $0 \le (At)^4 < 1$ and that
$s(t=0)=0$.
   
The metric function $s(t)$, and its derivatives, $\dot{s}(t)$ and
$\ddot{s}(t)$ are finite at $t=0$, then the Weyl scalar $\Psi_2$, Eq.
(\ref{Weylscalar}), and the invariants derived from it are regular at
$t=0$. This is an indication that singularity at $t=0$ is not a physical
one. However, $\ddot{s}(t)$ diverges as $At \to 1$ and it implies that the
polynomial invariants constructed from $\Psi_2$ diverge as well at $At=1$,
having there the possibility of a singularity. We shall analyze the
geodesics to figure out the singularity structure if it exists. Fig.
\ref{sdiag} displays the metric function $s(t)$, and its derivatives,
$\dot{s}(t)$ and $\ddot{s}(t)$.

The equations for the first derivatives of the coordinates with respect to
the affine parameter $\tau$ are equally valid in this case, just putting
$l=0$, $M=0$ and the metric function $s(t)$ from Eq. (\ref{s-diag}). If
the first derivatives are bounded it implies that geodesics are complete
and the field is nonsingular \cite{Arnold}; the first equations being,

\begin{eqnarray}
{y_{,\tau}}&=& \frac{K_2 \phi^2}{h},\nonumber\\
{x_{,\tau}}&=& \frac{K_1 \phi^2}{s},\nonumber\\
\end{eqnarray}
 
The term $\frac{ \phi^2}{s}$ that might diverge, is bounded in all the
range of $t$ and diverges at $t=0$ only when $\epsilon=0$, because

\begin{equation}
\lim_{t \to 0} \left(\frac{s}{\phi^2} \right)=- \frac{\epsilon}{A^2},
\end{equation}
in such a manner that geodesics ${x(\tau)}$ are unbounded for Bianchi
G$_3$I ($\epsilon=0$), since ${x_{,\tau}}$ diverges for a finite value of
the affine parameter. Even for the case $\epsilon=0$ if the constant
$K_1=0$ then the geodesics ${x(\tau)}$ are complete since in this case
${x(\tau)}=A \tau + B$, $A, B$ constants, and it is a complete curve.
 
The coefficients in equations for ${z_{,\tau}}$ and ${t_{,\tau}}$ are
bounded as can be seen from expressions (\ref{z,1})-(\ref{z,t,e=0})
putting $l=0$, $M=0$ . On the other side, in spite that

\begin{equation}
\lim_{at \to 1} (\ddot{s}) \to \infty,
\end{equation}
all the coefficients in the equations for $x^{\alpha}_{, \tau},
x^{\alpha}= (x, y,t,z)$ are bounded, in particular the term $\frac{
\phi^2}{s}$ is finite there,

\begin{equation}
\lim_{at \to 1} \left( \frac{s}{\phi^2} \right) =C_1A+
\frac{2 \sqrt{2}b^2}{3A^2}{\rm{K}}[\frac{1}{2}]-
\frac{2b^2}{3A^2}- \frac{\epsilon}{A^2}.
\end{equation} 

Then at ($at \to 1$) the geodesics are complete, showing that there 
is no singularity at that time.

In relation to the electromagnetic fields, it can be seen from Eqs.
(\ref{DBfields}) that when $l=0$ only electric or only magnetic field
persists (remind the electromagnetic duality, which is preserved in the
BI theory), for instance, $D(t)= \phi(t)^2$, $B=0$. But note that the
fields can not trespass the threshold of the maximum field $b$, then
$D(t)= A^2t^2 < b$ imposes an aditional restriction for the time
coordinate, it must be that $At<$Inf $\{ \sqrt{b}, 1 \}$.

\section{Singularity-free solutions with linear electromagnetic field}

The conserved quantities derived from the existence of the four Killing
vectors are independent of the linear or nonlinear electrodynamics. Then
the equations for the derivatives of the coordinates with respect to the
affine parameter can be analyzed to test completeness if instead of having
the BI field, we take the linear limit.

Expressions for the derivatives of the coordinates with respect to the
affine parameter, Eqs. (\ref{x,y})-(\ref{z,t,e=0}) are valid now with
$s(t)$ given by its linear limit, Eq. (\ref{sMax}). The possibility of
divergence appears again in the term $s/ \phi^2$ in the equation for $x_{,
\tau}$. We analyze the behavior of the term $s/ \phi^2$.  Integrating the
expression for $s(t)$ Eq. (\ref{sMax}) with $\phi=A \cos{t}$ we obtain:

\begin{equation}
s(t)= C_1 \phi^3 \dot{\phi}+\frac{C^2}{A^6} \left[{\phi^4-
\frac{5}{3}\phi^2 \dot{\phi}^2-\frac{A^2}{3} \dot{\phi}^2}\right]+
\frac{\epsilon}{A^4}[\phi^4- \phi^2 \dot{\phi}^2],
\label{sMax1}
\end{equation}
where $C^2=H^2+D^2$ is proportional to the energy density of the field
and the limits at $t \to 0$ and $t \to \pi/2$ are

\begin{eqnarray}
\lim_{t \to 0} s(t)&=&C^2/A^2+ \epsilon, \nonumber\\
\lim_{t \to \pi/2} s(t)&=&-C^2/3A^2,
\end{eqnarray}

From the previous expressions it is clear that $s(t)$ is not zero neither
at $t=0$ nor $t=\pi/2$, being the presence of the electromagnetic field
the term that allows the completeness of the geodesics; therefore the
coefficients in the equation for $x_{, \tau}$ are bounded and consequently
the geodesic $x( \tau)$ is complete. This proves that for the solution of
the Einstein-Maxwell equations given by the line element (\ref{metric})
with $s(t)$ in Eq. (\ref{sMax1}) there are singularity-free spacetimes.

Moreover, for the case $l=0$ we also have singularity-free solutions since
the term $s/ \phi^2$ is not zero neither at $t=0$ not at $At=1$; in this
case $\phi=A t$ and

\begin{equation}
s(t)= C_1 A^4t^3-C^2- \epsilon t^2,
\label{sMax2}
\end{equation}

and the limits

\begin{eqnarray}
\lim_{t \to 0} s(t)&=&-C^2, \nonumber\\
\lim_{at \to 1} s(t)&=&C_1A-C^2- \epsilon/A^2,
\end{eqnarray}
  
Showing that the geodesics are complete, since the coefficients in the
equation for $x_{, \tau}$ are bounded. Summarizing, there are
singularity-free solutions with constant electromagnetic field for
spacetimes given by the metric (\ref{metric}) and the metric function
$s(t)$ given by (\ref{sMax1}) if $l \ne 0$ or by (\ref{sMax2}) if $l=0$.
The electromagnetic field is given by Eqs. (\ref{linearfield}). We also
note that for these singularity-free solutions the electromagnetic field
tensor does not violate SEC.


\section{Conclusions}
 
We investigated spacetimes with a four-dimensional group of
isometries with a three-dimensional subgroup acting transitively on
spacelike hypersurfaces, i. e. spatially homogeneous spacetimes
anisotropic in one spatial direction. We address the study of geodesic
completeness
for solutions to nonlinear electromagnetic field coupled with
Einstein equations for cases of Bianchi $G_{3}I$, $G_{3}II$,
$G_{3}III$ $G_{3}VIII$, $G_{3}IX$ and Kantowski-Sachs.

Two families are distinguished: Bianchi $G_{3}I$, $G_{3}III$ and
Kantowski-Sachs on one side and Bianchi $G_{3}II$, $G_{3}VIII$ and
$G_{3}IX$ on the other.
 
Bianchi $G_{3}I$, $G_{3}III$ and Kantowski-Sachs only admit one
component of the electromagnetic field, electric or magnetic (not both,
interchanging roles by a duality rotation).  These spaces show divergence
in their invariants 
at a finite time, however, geodesics are complete.
At $t=0$ it is shown that geodesic uncompleteness appears for the case
$\epsilon=0$.

Types VIII and IX are geodesically complete and as such they are
singularity-free. The space VIII resembles the ones studied by Siklos
\cite{Siklos} as extensions of Taub-NUT metrics. When Bianchi G$_3$IX is
considered as having the topology $RXS^3$ it presents imprisoned
incompleteness, described by Hawking (see paragraph 8.5 in \cite{Hawking})
for Taub-NUT spaces. This pathological behavior arises essentially as a
consequence of assuming compactness for Bianchi type IX.

Our results are not in contradiction with the assertion by Hawking about
the necessity of singularity in non-empty spatially homogeneous models,
since the timelike convergence condition fails to hold in nonlinear
electrodynamics. The incompleteness due to the presence of
imprisoned geodesics in Bianchi G$_3$IX seems to be inconsistent with the
affirmation that: {\it in the Taub-NUT case the incompleteness of
geodesics leads to a real singularity once any amount of matter is added}
(see for instance \cite{Ryan}).  In this paper we have presented a
counterexample for a space of Taub-NUT type with imprisoned geodesics but
there is no contradiction since we have considered as matter a nonlinear
electromagnetic field and the original assertion was made thinking of
matter as a perfect fluid.
  
Finally we remark that in the linear limit (Einstein-Maxwell) the
solutions are singularity-free for all the Bianchi cases included, being
clearly in this case the presence of the field that makes the geodesics
complete. In the nonlinear case the determinant factor to avoid the
singularity is the curvature parameter, existing singular solutions if
the curvature parameter is zero ($\epsilon=0$).


\begin{acknowledgments}
R. G-S. acknowledges partial support from
project UAEM 1941/2004.
\end{acknowledgments}


\end{document}